\documentclass[11pt]{article}

\usepackage{graphicx,amsfonts,amssymb}
\usepackage[mathscr]{euscript}
\usepackage{graphics}
\usepackage{epsfig,psfrag,amsfonts,amssymb,amsmath}
\begin{document} 
\makeatletter
\@addtoreset{equation}{section}
\makeatother
\renewcommand{\theequation}{\thesection.\arabic{equation}}
\baselineskip 15pt

\title{\bf An optimal entropic uncertainty relation in a two-dimensional 
 Hilbert space\footnote{Work supported in part by Istituto Nazionale di Fisica 
 Nucleare, Sezione di Trieste, Italy}}
\author{GianCarlo Ghirardi\footnote{e-mail: ghirardi@ts.infn.it}\\
{\small Department of Theoretical Physics of the University of Trieste, and}\\
{\small International Centre for Theoretical Physics ``Abdus Salam'', and}\\
{\small Istituto Nazionale di Fisica Nucleare, Sezione di Trieste, Italy}\\
and \\
\\ Luca Marinatto\footnote{e-mail: marinatto@ts.infn.it}\\
{\small International Centre for Theoretical Physics ``Abdus Salam'', and}\\
{\small Istituto Nazionale di Fisica Nucleare, Sezione di Trieste, Italy}\\
and \\
\\ Raffaele Romano\footnote{e-mail: rromano@ts.infn.it }\\
{\small Department of Theoretical Physics of the University of Trieste, and}\\
{\small Istituto Nazionale di Fisica Nucleare, Sezione di Trieste, Italy}}

\title{\vspace{.5cm}
\bf {\Large An optimal entropic uncertainty relation in a two-dimensional 
 Hilbert space\footnote{Work supported in part by Istituto Nazionale di Fisica
 Nucleare, Sezione di Trieste, Italy}}}

\date{}
\maketitle

\begin{abstract}
We derive an optimal entropic uncertainty relation for an arbitrary pair of
 observables in a two-dimensional Hilbert space. Such a result, for the simple
 case we are considering, definitively improves all the entropic 
 uncertainty relations which have appeared in the literature.\\

  Key words: Entropy; Uncertainty relation; Complementary observables.\\

  PACS: 0.3.65.Bz

\end{abstract}


\section{Introduction}

The uncertainty principle of quantum mechanics is expressed by the
 well-known Robertson relation~\cite{rob}:
\begin{equation}
\label{intro1}
\Delta_{\psi}A \Delta_{\psi}B \geq \frac{1}{2}\vert \langle \psi \vert
\:[A,B]
\vert \:\psi \rangle \vert
\end{equation}
where $\Delta_{\psi}A$ and $ \Delta_{\psi}B$ represent the variances
 of the observables $A$ and $B$ when the state of the system is $\vert \psi
\rangle$.
This inequality, when the expectation value of the commutator $[A,B]$ 
 does not vanish, expresses the intrinsic quantum mechanical limitation on the 
 possibility of preparing homogeneous quantum ensembles with arbitrarily 
 narrow variances for the involved observables and gives, in general,
 a physically useful information about the considered observables for the
 pure case associated to the state $\vert \psi \rangle$.
In a Letter of various years ago, Deutsch~\cite{deu} moved a compelling 
 criticism about the
 inadequacy of such inequality on the ground that a true indeterminacy
 relation should not exhibit a dependence on the state vector on its
 right-hand side, as it happens with Eq.~(\ref{intro1}).
Indeed, according to the author, such an hypothetical relation should be an 
 inequality whose left-hand side quantifies, in a way to be defined 
 appropriately, the
 uncertainty in the results of measurement processes of a pair of observables,
 while the right-hand side should contain a fixed and irreducible lower
 bound.
This kind of mathematical expression cannot be generally obtained when dealing
 with a relation displaying the form of Eq.~(\ref{intro1}).
In fact, given a pair of non-commuting observables $A$ and $B$, 
 belonging to an arbitrary Hilbert space $\cal{H}$, it is easy to show that the 
 quantity $\Delta_{\psi} A \Delta_{\psi} B$ can either vanish or become
 arbitrarily close to zero if at least one of the two observables is bounded.
The proof of this goes as follows: suppose $B$ is the bounded observable
 and suppose $A$ possesses a discrete eigenvalue.
In this case the variance of $A$ becomes null in correspondence of the 
 proper eigenvector associated to the discrete eigenvalue and the 
 indeterminacy relation assumes the desired, but absolutely trivial, form
\begin{equation}
\label{intro2}
 \Delta_{\psi} A \Delta_{\psi} B \geq 0\:\:\:\:\:\:\: \forall\:
 \vert \psi \rangle\:\in\:{\cal{H}},
\end{equation}  
since $\Delta_{\psi} B$ is always finite for bounded $B$. 
A similar result holds when the observable $A$ has a purely continuous 
 spectrum: also in this case  the variance $\Delta_{\psi} A$ can be made 
 arbitrarily close to zero ($\forall \epsilon >0,\:\:\exists \:\vert \psi
 \rangle \in {\cal H}: \Delta_{\psi} A\leq \epsilon$), while the 
 other variance remains bounded. 
On the contrary, a non-trivial uncertainty relation (of the kind we are
 searching for) can be written whenever the
 commutator of $A$ and $B$ is equal to a multiple of the identity, 
 a case implying that both operators are unbounded.
The paradigmatic example is represented by the pair of  position $Q$ and 
 momentum $P$ operators for which $[Q,P]=i\hbar$: in this case the usual
 uncertainty relation $\Delta_{\psi}Q\Delta_{\psi}P\geq \hbar/2$
 is obtained.
This formula is significant since it exhibits a non-zero irreducible
 lower bound $\hbar/2$ constraining the possible values of the two variances.  
 
So, if one pretends that the right-hand side of an indeterminacy relation
 of the type~(\ref{intro1}) does not depend on the chosen state $\vert \psi
 \rangle$, one unavoidably ends up with the trivial result~(\ref{intro2}),   
 for every pair of observables in any Hilbert space when at least one of them 
 is bounded.
 
In order to overcome this problem Deutsch proposed
 Shannon entropy as an optimal measure of the amount of uncertainty which
 should be naturally connected with the measurement process of a pair
 of observables.
Given an observable $A$ of a Hilbert space $\cal{H}$ with purely discrete 
 spectrum $\left\{ a_{i} \right\}$,
 the Shannon entropy of $A$ in the state $\vert \psi \rangle$ is defined
 as the quantity
\begin{equation}
\label{intro3}
S_{\psi}(A) \equiv - \sum_{i} p_{i} \log p_{i}
\end{equation}
in terms of the probabilities $p_{i}$ of getting the eigenvalue $a_{i}$
 in a measurement of the observable $A$ in the given state $\vert \psi \rangle$.

Now, in the particular case in which the couple of observables $A$ and
 $B$ of $\cal{H}$ have a non-degenerate spectrum, calling 
 $\vert a_{i} \rangle$ and  $\vert b_{j} \rangle$ the (unique)
 eigenvectors associated to the eigenvalues $a_{i}$ and $b_{j}$
 respectively, the following relevant entropic uncertainty relation holds:
\begin{equation}
\label{intro4}
S_{\psi}(A) + S_{\psi}(B) \geq 2 \log\Big(\frac{2}{1+ \sup_{ij} | \langle
 a_{i} | b_{j} \rangle |}\Big)\:\:\:\:\:\:\: \forall\: \vert \psi \rangle
\in \:{\cal{H}}.
\end{equation}
This inequality, originally derived by Deutsch~\cite{deu} under rather
 restrictive assumptions, has been shown later to be a particular case of a 
 more general formula involving totally arbitrary observables~\cite{partovi} 
 and has the appealing
 advantage of displaying a right-hand side which is independent of the
 state $\vert \psi \rangle$.
Moreover it gives a non-trivial information, i.e. a strictly positive lower
 bound, concerning the sum of the uncertainties associated to measurement
 outcomes when the observables $A$ and $B$ do not have any common eigenvector.
A further improvement of the previous formula, in the case of a 
 finite-dimensional Hilbert space, has been subsequently obtained  
 in Ref.~\cite{uffink}, where it has been proved that 
\begin{equation}
\label{intro4.5}
S_{\psi}(A) + S_{\psi}(B) \geq -2 \log\Big( \max_{ij} | \langle
 a_{i} | b_{j} \rangle |\Big)\:\:\:\:\:\:\: \forall\: \vert \psi \rangle
\in \:{\cal{H}}
\end{equation}
However, the quantities appearing at the right-hand side of Eqs.~(\ref{intro4})
 and~(\ref{intro4.5}) are in general not optimal.
The optimal lower bound for two given observables $A$ and $B$ can only be  
 found by calculating explicitly the 
 minimum of $S_{\psi}(A)+S_{\psi}(B)$ over all the normalized state
 vectors $\vert \psi \rangle\in \cal{H}$.

In this Letter we show how to determine exactly such an (optimal)
 lower bound in the particular case of a two-dimensional Hilbert
 space (i.e., when $n\!=\!2$).
Such a value significantly improves the lower bounds given by 
 Eqs.~(\ref{intro4}) and~(\ref{intro4.5}).


\section{Optimal entropic uncertainty relation when $n=2$}

Let us consider two arbitrary Hermitian operators $A=( \alpha_{1}I +
 \beta_{1}\vec{\sigma}\cdot \vec{m})$ and $B=(\alpha_{2} I +
 \beta_{2}\vec{\sigma}\cdot \vec{n})$ of ${\cal{H}}=\mathbb{C}^{2}$, where 
 $(\alpha_{i},\beta_{i})$ are
 real numbers, $\vec{\sigma}=(\sigma_{x}, \sigma_{y}, \sigma_{z})$ are the
 Pauli matrices and $(\vec{m},\vec{n})$ are two unit vectors
 in the three-dimensional Euclidean space.
According to what has been said previously, our purpose consists in determining
 the following quantity
\begin{equation}
\label{optimal1}
\min_{\vert \psi \rangle}\: [\: S_{\psi}(A) + S_{\psi}(B)\: ],
\end{equation}
where the minimum is calculated over the set of all normalized states
 $\vert \psi \rangle\! \in\! \mathbb{C}^{2}$, for fixed $\alpha_{i},
 \beta_{i}, \vec{m}, \vec{n}$.

In order to simplify the calculations of the quantities $S_{\psi}(A)$ and
 $S_{\psi}(B)$, we begin by reducing the number of the involved parameters.
To this purpose we note that it is possible to restrict our attention to
 the class of Hermitian operators of the form $A=\vec{\sigma}\cdot\vec{m}$
 and $B=\vec{\sigma}\cdot\vec{n}$ only.
In fact the Shannon entropy $S_{\psi}(A)$ (an equivalent consideration
 holds for $S_{\psi}(B)\:$) does not depend on the eigenvalues of the involved 
 observable but only on the scalar products of its eigenstates
 $\left\{\: \vert \uparrow \vec{m} \rangle, \vert \downarrow \vec{m} \rangle\:
 \right\}$ with the state $\vert \psi \rangle$.
Since the eigenstates of $A= \alpha_{1}I +\beta_{1} \vec{\sigma}\cdot\vec{m}$ 
 are the same as those of the simpler operator 
 $A=\vec{\sigma}\cdot\vec{m}$, our simplification does not affect the
 final result.
 
Accordingly, the entropy $S_{\psi}(A)$ of Eq.~(\ref{intro3})
 takes the form
\begin{equation}
\label{optimal2}
 S_{\psi}(A)= - \vert \langle \vec{m}\uparrow \vert \psi\rangle \vert^{2}
 \log  \vert \langle \vec{m} \uparrow \vert \psi \rangle\vert^{2} -
 \vert \langle \vec{m} \downarrow \vert \psi \rangle\vert^{2}
 \log  \vert \langle \vec{m} \downarrow \vert \psi \rangle\vert^{2}
\end{equation}
and the corresponding formula for $S_{\psi}(B)$ is obtained by replacing
 $\vec{m}$ with $\vec{n}$.

The two-dimensional Hilbert space we are dealing with exhibits two
 remarkable geometrical properties which will be of help in
 simplifying the expression~(\ref{optimal2}): 
 i) every normalized state $\vert \psi \rangle \in \mathbb{C}^{2}$ can be
 associated to a unit vector $\vec{k}$ in the three-dimensional Euclidean space
 $\mathbb{R}^{3}$ (all these vectors forming the so-called Bloch sphere)
 by requiring that $\vert \psi \rangle$ is the eigenstate of the observable
 $\vec{\sigma}\cdot\vec{k}$ pertaining to the eigenvalue $+1$;
 ii) scalar products of state vectors belonging to $\mathbb{C}^{2}$ are
 simply related to the Euclidean scalar product of their corresponding
 three-dimensional vectors, according to $\vert\langle \vec{m} \uparrow
 \vert \vec{k} \uparrow \rangle \vert^{2}= \frac{1}{2}(1 + \vec{m}
 \cdot \vec{k})$ and $\vert\langle \vec{m} \downarrow
 \vert \vec{k} \uparrow \rangle \vert^{2}= \frac{1}{2}(1 - \vec{m}
 \cdot \vec{k})$.
 
Owing to these nice features, the sum $S_{\psi}(A) + S_{\psi}(B)$
 displays the following appealing form:
\[
S_{\psi}(A) + S_{\psi}(B)= - \frac{1}{2}(1- \vec{m}\cdot\vec{k})
\log  \frac{1}{2}(1- \vec{m}\cdot\vec{k}) -
 \frac{1}{2}(1+ \vec{m}\cdot\vec{k})
\log  \frac{1}{2}(1+ \vec{m}\cdot\vec{k})
\]
\begin{equation}
\label{optimal3}
- \frac{1}{2}(1- \vec{n}\cdot\vec{k})
\log  \frac{1}{2}(1- \vec{n}\cdot\vec{k}) -
 \frac{1}{2}(1+ \vec{n}\cdot\vec{k})
\log  \frac{1}{2}(1+ \vec{n}\cdot\vec{k}).
\end{equation}
Our problem consists in finding the minimum value of the above
 expression over all possible orientations of the (three-dimensional)
 unit vector $\vec{k}$ for fixed $\vec{m}$ and $\vec{n}$.
So, the quest for the minimum of the sum of the Shannon entropies of two
 arbitrary observables $A$ and $B$ has been turned into the geometrical
 problem of finding the Euclidean vector $\vec{k}$ which minimizes the 
 quantity~(\ref{optimal3}) for the two fixed spatial directions.

The problem can be further simplified by observing that the minimum
 of~(\ref{optimal3}) is attained when $\vec{k}$ lies on the plane $\tau$,
 determined by the vectors $\vec{m}$ and $\vec{n}$.
In order to prove that let us consider an arbitrary plane $\tau_{\perp}$
 which is orthogonal to $\tau$, and let us take into account the function, 
 appearing in Eq.~(\ref{optimal3}), $H_{0}(x)= -\frac{1}{2}(1-x)\log
 \frac{1}{2}(1-x)- \frac{1}{2}(1+x)\log \frac{1}{2}(1+x)$ for $x \in [0,1]$.
Such a function is monotonically decreasing from its maximum value $\log 2$ to 
 $0$. 
Accordingly $H_{0}(x_{0})\leq H_{0}(x)$ for $x_{0}\geq x$.
Let us identify  now the variable $x$ with $\vert \vec{m}\cdot \vec{k}\vert$
 for $\vec{k}$ any unit vector belonging to $\tau_{\perp}$.
If we choose as $\vec{k}_{0}$ one of the two unit vectors of  $\tau_{\perp}$
 belonging to $\tau$, we obviously have that $\vert \vec{m}\cdot \vec{k}\vert
 \leq \vert \vec{m}\cdot \vec{k}_{0}\vert\:\:\forall \vec{k}\in \tau_{\perp}$,
 the equality sign holding when $\vec{k}=\pm \vec{k}_{0}$.
There follows that $H_{0}( \vert \vec{m}\cdot \vec{k}_{0}\vert) \leq
 H_{0}( \vert \vec{m}\cdot \vec{k}\vert)$ (an analogous consideration holds 
 when $\vec{m}$ is replaced with $\vec{n}$) which implies that to find the
 minimum of~(\ref{optimal3}) we can confine our considerations to unit
 vectors belonging to the plane $\tau$.
 
Having restricted our attention to an expression involving only vectors
 lying on the plane identified by the vectors $\vec{m}$ and $\vec{n}$, 
 we introduce the (fixed) angle $\alpha$ between these two vectors
 and the (freely variable) angle $\theta$ formed by $\vec{m}$ and $\vec{k}$. 
Due to symmetry considerations the angle $\alpha$
 can be chosen to belong to the interval $[0,\pi)$.
 
In terms of these angles, we can now tackle the problem of finding the minimum
 of the quantity of interest 
\[
S_{\psi}(A) + S_{\psi}(B)= - \frac{(1- \cos\theta)}{2}
 \log  \frac{(1- \cos\theta)}{2} -  \frac{(1+ \cos\theta)}{2}
 \log  \frac{(1+ \cos\theta)}{2}
\]
\begin{equation}
 \label{optimal4}
 - \frac{(1-\cos(\alpha-\theta))}{2} \log  \frac{(1-
 \cos(\alpha-\theta))}{2}
 - \frac{(1+ \cos(\alpha-\theta))}{2}
 \log  \frac{(1+ \cos(\alpha-\theta))}{2},
\end{equation}
when $\theta$ runs over the interval $[0,2\pi)$, for a fixed 
 $\alpha\in [0,\pi)$.
The above expression depends smoothly on $\theta$ (and it does not display
 any singularity) and the number of its minima depends on the value of 
 $\alpha$.
In fact, from numerical plots of~(\ref{optimal4}), one easily notices that
 there is a critical value of $\alpha$, which we denote as $\bar{\alpha}$,
 in correspondence of which the number of absolute minima of 
 $S_{\psi}(A) + S_{\psi}(B)$ within $[0,2\pi)$, changes from two to four 
 (we will denote this phenomenon as parametric bifurcation).

More precisely an analytical expression for the minimum value 
 of~(\ref{optimal4}), in terms of the angle $\alpha$,
 can be given when $\alpha \in [0,\bar{\alpha}] \cup
 [\pi -\bar{\alpha}, \pi)$, such a minimum being attained for two well-defined 
 directions.
On the contrary, when $\alpha \in (\bar{\alpha},\pi-\bar{\alpha})$ there are
 four minima and the minimum value cannot be given in terms of
 elementary functions of $\alpha$ but can only be determined by means of 
 numerical calculations.
Let us now analyze in detail and separately all possible cases.

For $\alpha\in [0,\bar{\alpha}]$ the vectors $\vec{k}$ which
 minimize~(\ref{optimal4}) lay half-away between the directions $\vec{m}$ and 
 $\vec{n}$, that is the minima are
 located at $\theta = \alpha/2 $ and $\theta = \pi + \alpha/2$.
Such a result is obtained by calculating the first and second order
 derivatives of~(\ref{optimal4}) with respect to $\theta$: only when $\theta =
 \alpha/2 $ and $\theta = \pi + \alpha/2$ the first derivative vanishes 
 while the second turns out to be always positive.
 
In this case we can easily obtain an analytical expression for the optimal
 entropic uncertainty relation, depending only on the angle $\alpha$ 
 specified by the two observables:

\begin{equation}
\label{optimal5}
 S_{\psi}(A) + S_{\psi}(B) \geq
 - (1- \cos\frac{\alpha}{2})
\log  \frac{1}{2}(1- \cos\frac{\alpha}{2}) -  (1+ \cos\frac{\alpha}{2})
\log  \frac{1}{2}(1+ \cos\frac{\alpha}{2})\:\:\:\:\:\forall\:
\vert\psi\rangle \in {\cal H}.
\end{equation}

For $\alpha \in (\bar{\alpha}, \pi/2)$ the bifurcation phenomenon occurs 
 and four mimima appear.
As $\alpha$ grows, two minima spring symmetrically out of $\theta = \alpha/2$ 
 and move towards the vectors $\vec{m}$ and $\vec{n}$, while the other two are 
 symmetrically placed with respect to the opposite direction $\theta =
 \pi +\alpha/2$ and move towards $-\vec{m}$ and $-\vec{n}$.
In this case an analytical expression for the entropic uncertainty principle
 in terms of elementary functions cannot be given and we need to resort
 to direct numerical calculations for obtaining the value of the minimum 
 for every fixed value of $\alpha$ (see Fig.~\ref{fig1}).
 
When $\alpha= \pi/2$ we are considering a couple of complementary 
 observables~\cite{kraus}. 
In a two-dimensional Hilbert space, complementary observables are represented
 by spin components along orthogonal axis, and their entropic uncertainty
 relation reduces to:
\begin{equation}
\label{optimal7}
  S_{\psi}(A) + S_{\psi}(B) \geq \log 2 \:\:\:\:\:\:\:\:
  \forall\:\vert\psi\rangle\in {\cal H}.
\end{equation}
For $\alpha \in (\pi/2, \pi-\bar{\alpha})$ there exist again four vectors 
 $\vec{k}$ which minimize~(\ref{optimal4}): two of them are located 
 symmetrically around the direction $\theta = (\pi +\alpha)/2$ and move 
 towards it from $\vec{n}$ and $-\vec{m}$; the other two are located
 around $\theta = (3\pi +\alpha)/2$ moving towards it from $-\vec{n}$ and
 $\vec{m}$.
In this case also, we have to resort to numerical calculations for
 obtaining an exact value of the minimum of $S_{\psi}(A) + S_{\psi}(B)$ and
 its values are plotted in Fig.~\ref{fig1}.
 
Finally, when $\alpha \in [\pi-\bar{\alpha},\pi)$, the bifurcation disappears
 and only two minima remain corresponding to vectors $\vec{k}$ directed along 
 $\theta=(\pi +\alpha)/2$ and $\theta = (3\pi +\alpha)/2$. 
An analytical expression for the entropic uncertainty relation can be
 given in terms of the angle $\alpha$ in this situation too:   
\begin{equation}
\label{optimal8}
 S_{\psi}(A) + S_{\psi}(B) \geq - (1+ \sin\frac{\alpha}{2})
\log  \frac{1}{2}(1+\sin\frac{\alpha}{2}) -  (1- \sin\frac{\alpha}{2})
\log  \frac{1}{2}(1- \sin\frac{\alpha}{2})\:\:\:\:\:\forall\:
\vert\psi\rangle \in {\cal H}. 
\end{equation}
An exact plot displaying the minimum value of $S_{\psi}(A)+S_{\psi}(B)$ with 
 respect to the variable $\alpha$ can be finally obtained by plotting 
 Eqs.~({\ref{optimal5}) and~({\ref{optimal8}) together with the numerical 
 values obtained when $\alpha \in (\bar{\alpha}, \pi-\bar{\alpha})$.
\begin{figure}[th!]
\centerline{\epsfig{file=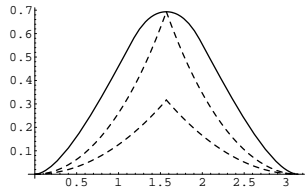, width=4cm}} \vspace*{8pt}
\caption{\label{fig1}Minimum value of $S_{\psi}(A) + S_{\psi}(B)$ with
respect to angle $\alpha\in [0,\pi)$ in solid line, and non-optimal
 estimates~(\ref{intro4}) [the bottom one] and~(\ref{intro4.5}) [the top one] 
 in dashed lines.}
\end{figure}

The dashed curves of Fig.~\ref{fig1} represent the unotpimal estimates 
 for the entropic uncertainty relations~(\ref{intro4}) and~(\ref{intro4.5}): 
 the plot clearly shows how much our result, represented by the curve
 in solid line, outperforms the currently known lower 
 bounds, when $\alpha\neq \pi/2$. 
Such curves, when $n\!=\!2$, can be easily proved to be equal to
 $S_{\psi}(A) + S_{\psi}(B)\geq -2\log\frac{1}{2}(1+c)$ and 
 $S_{\psi}(A) + S_{\psi}(B)\geq -2\log c$ respectively, where
 $c=\cos\frac{\alpha}{2}$ when $\alpha \in [0,\pi/2)$ and 
 $c=\sin\frac{\alpha}{2}$ when $\alpha \in [\pi/2,\pi)$.
 
Fig.~\ref{fig2} shows instead the angular position $\theta\in [0,2\pi)$ 
 (on the vertical axis) of all the possible minima of 
 $S_{\psi}(A) + S_{\psi}(B)$ 
 with respect to the variable $\alpha \in [0,\pi)$ (on the horizontal axis): 
 the peculiar phenomenon of the bifurcation is clearly visible in 
 correspondence of the critical points $\bar{\alpha}$ and $\pi-\bar{\alpha}$
 (in dashed line).

\begin{figure}[th!]
\centerline{\epsfig{file=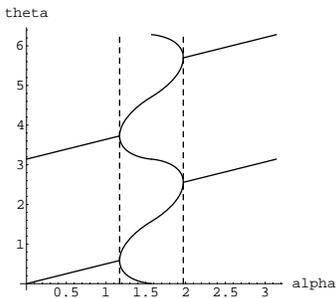, width=4.5cm}}  \vspace*{8pt}
\caption{\label{fig2}Angular position of the minima of
$S_{\psi}(A) + S_{\psi}(B)$ with respect to angle $\alpha\in [0,\pi)$.}
\end{figure}

It remains to explain how to determine the
 angular value $\bar{\alpha}$ for which the bifurcation phenomenon appears.
From an analysis of~(\ref{optimal4}) when $\alpha \in [0,\pi/2]$, 
 we notice that the angle $\theta = \alpha/2$ is always an extremal point
 for the sum of the two entropies (since the first derivative
 of~(\ref{optimal4}) vanishes in it), passing from a (relative) minimum
 to a (relative) maximum point in correspondence of the critical value 
 $\bar{\alpha}$.  
Therefore, if we take the second derivative of~(\ref{optimal4})
 with respect to the variable $\theta$ and we evaluate it for $\theta =
 \alpha/2$, the function we obtain must change sign exactly for
 the critical value $\alpha=\bar{\alpha}$. 
Therefore the desired critical angle $\bar{\alpha}$ turns out to be the 
 (unique) solution of the following equation within the range $[0,\pi/2]$:
\begin{equation}
\label{optimal9}
\frac{\partial^{2}}{\partial\theta^{2}}[S_{\psi}(A) + S_{\psi}(B)]
\Big|_{\theta = \alpha/2}=-\cos\frac{\alpha}{2}\cdot
\log\Big[\Big(\tan\frac{\alpha}{4}\Big)^2\Big]
 - 2=0\:.
 \end{equation}
From a numerical analysis, we have obtained the 
 approximated value $\bar{\alpha}\simeq 1.17056$ for the critical angle.
 

\section{Conclusions}

The entropic uncertainty relations for a pair of observables in a 
 finite-dimensional Hilbert space constitute an appealing measure of the 
 degree of uncertainty for measurement outcomes. 
In the particular case of a two-dimensional Hilbert space, we have been
 able to determine (in part analitycally and in part numerically) 
 an optimal entropic uncertainty relation for spin-observables which improves 
 the known lower bounds.
 
Our procedure works easily in this case, since we have been able to reduce 
 our problem to a simple geometrical one by resorting to the well-known 
 correspondence between state vectors and points of a unit sphere within the 
 three-dimensional Euclidean space.  
An analogue result for higher dimensional spaces is still lacking.



\end{document}